\documentclass
[preprintnumbers,noshowpacs,nobibnotes,titlepage,prb,a4paper,12pt,onecolumn]{revtex4}%
\usepackage{graphicx}
\usepackage{amsmath}
\usepackage{amsfonts}
\usepackage{amssymb}%
\providecommand{\U}[1]{\protect\rule{.1in}{.1in}}
\begin{document}
\title{Novel Self-Sustained Modulation in Superconducting Stripline Resonators}
\author{Eran Segev$^{\ast}$}
\author{Baleegh Abdo}
\author{Oleg Shtempluck}
\author{Eyal Buks}
\affiliation{Microelectronics Research Center, Department of Electrical Engineering,
Technion, Haifa 32000, Israel}
\maketitle

\textbf{Nonlinear effects in superconductors are important for both basic
science and technology. A strong nonlinearity may be exploited to demonstrate
some important quantum phenomena in the microwave region, such as quantum
squeezing \cite{Sqz_Movshovich90, Squeezing_Yurke05, SQZ_Buks05} and
experimental observation of the so called dynamical Casimir effect
\cite{segev06e}; Whereas technologically, these effects may allow some
intriguing applications such as bifurcation amplifiers for quantum
measurements \cite{BifAmp_siddiqi06,BifAmp_Wiesenfeld86} and resonant readout
of qubits \cite{ResRead_Janice06}. In this work we study the response of a
superconducting (SC) microwave stripline resonator to a monochromatic injected
signal. We find that there is a certain range of driving parameters, in which
a novel nonlinear phenomenon immerges, and self-sustained modulation (SM) of
the reflected power off the resonator is generated by the resonator. That is,
the resonator undergoes limited cycle oscillations, ranging between several to
tens of megahertz. A theoretical model which attributes the SM to a thermal
instability yields a good agreement with the experimental results. A similar
phenomenon was briefly reported in the 60's
\cite{SM_Clorfeine64,DAiello66,SM_Peskovatskii67,SM_ERU70} in dielectric
resonators which were partially coated by a SC film, but it was not thoroughly
investigated and therefore its significance was somewhat overlooked. This
phenomenon is of a significant importance as it introduces an extreme
nonlinear mechanism, which is by far stronger than any other nonlinearity
observed before in SC resonators \cite{segev06d}. It results in high
intermodulation gain, substantial noise squeezing, period doubling\textbf{ of
various orders \cite{segev06d}, and strong coupling between resonance modes
(see supplementary figure 1)}.}%

\begin{figure}
[h]
\begin{center}
\includegraphics[
height=1.3159in,
width=3.4014in
]%
{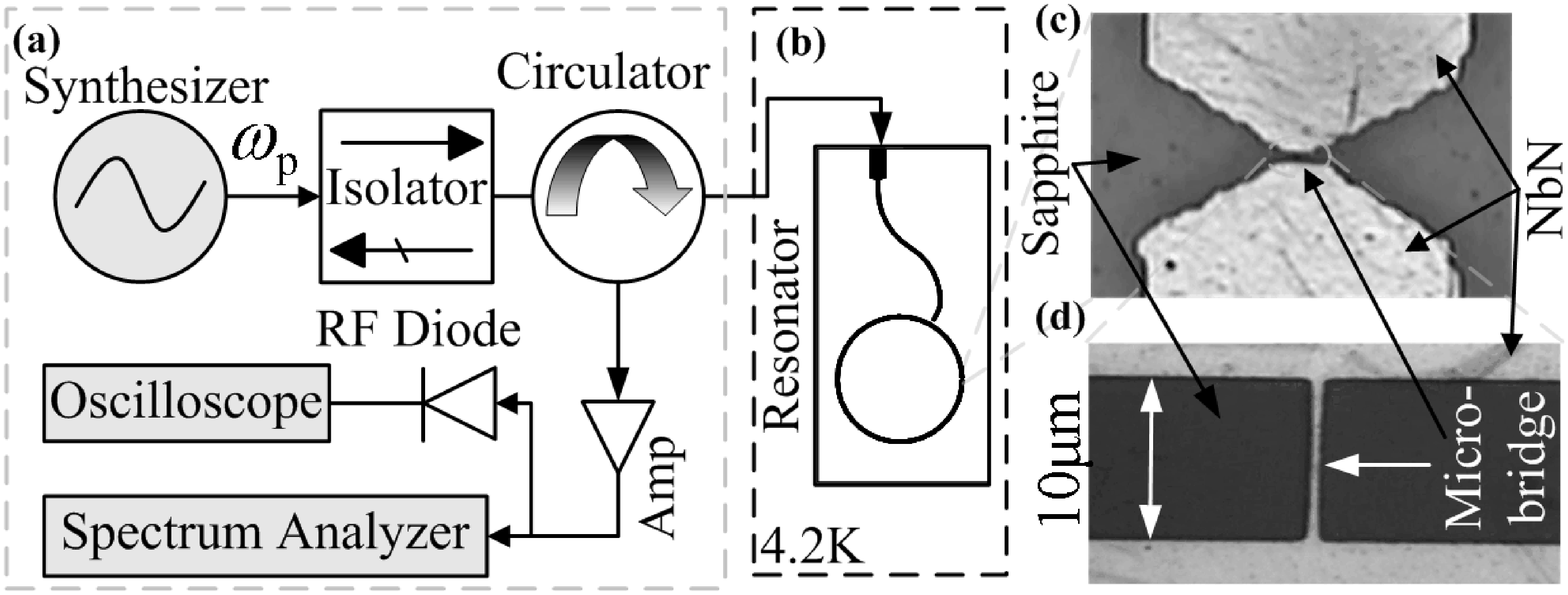}%
\caption{$(\mathrm{a})$ SM measurement setup. $(\mathrm{b})$ Schematic layout
of the device. The resonator is designed as a stripline ring, having a
characteristic impedance of $50\operatorname{\Omega }$. It is composed of
200nm thick NbN deposited on a Sapphire wafer. A weakly coupled feedline is
employed for delivering the input and output signals. An optical microscope
image of the ring resonator section at which the microbridge is integrated is
seen in panel $(\mathrm{c})$, whereas panel $(\mathrm{d})$ shows the
microbridge, whose dimensions are $1\times10\operatorname{\mu m}^{2}$.}%
\label{ExpSetup_P_Device}%
\end{center}
\end{figure}
Our device (Fig. \ref{ExpSetup_P_Device} $(\mathrm{b})-(\mathrm{d})$)
integrates a narrow microbridge into a SC stripline ring resonator. The
impedance of the microbridge strongly affects the normal modes of the
resonator and thus the resonance frequencies can be tuned by either internal
(Joule self heating) or external (infrared illumination \cite{Segev06a})
perturbations \cite{supRes_Saeedkia05}. Further design considerations,
fabrication details as well as normal modes calculation can be found elsewhere
\cite{Segev06a}. The experiments are performed using the setup described in
Fig. \ref{ExpSetup_P_Device}$(\mathrm{a})$. We stimulate the resonator with a
monochromatic tone at an angular frequency $\omega_{\mathrm{p}}$, and measure
the reflected power off the resonator by a spectrum analyzer in the frequency
domain and an oscilloscope in the time domain. Measurements are carried out
while the device is fully immersed in liquid Helium.%

\begin{figure*}
\centering
\[%
\begin{array}
[c]{cc}%
\text{%
\raisebox{-0cm}{\parbox[b]{5.5882cm}{\begin{center}
\includegraphics[
height=4.751cm,
width=5.5882cm
]%
{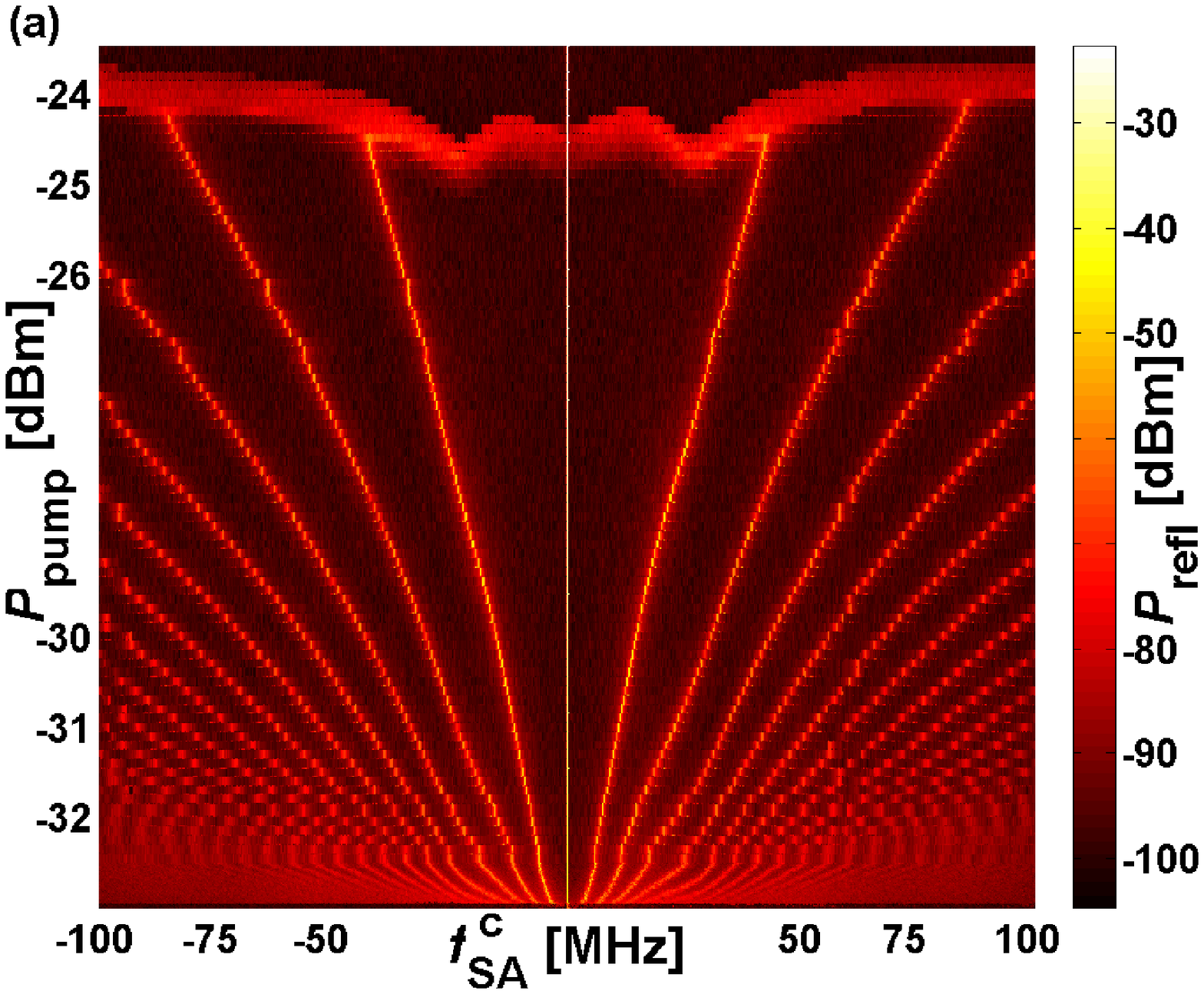}%
\\
{}%
\end{center}}}%
} & \text{%
\raisebox{0.0211cm}{\parbox[b]{6.1196cm}{\begin{center}
\includegraphics[
height=4.6667cm,
width=6.1196cm
]%
{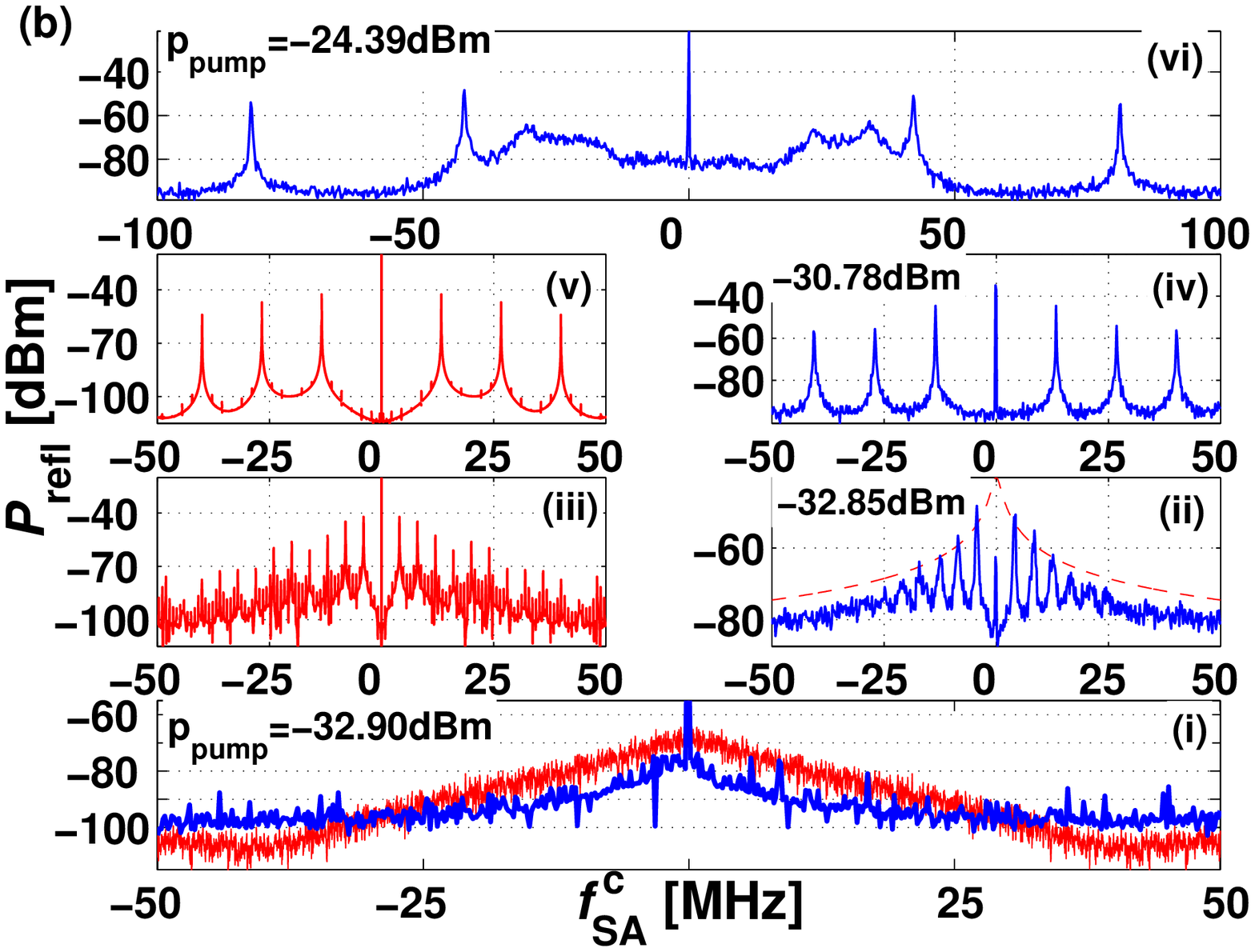}%
\\
{}%
\end{center}}}%
}%
\end{array}
\]
%

\caption[h]
{Typical experimental results of the SM phenomenon in the frequency domain.
Panel $(\mathrm{a})$ plots a color map of the reflected power $P_{\mathrm
{refl}}$ as a function of the pump power $P_{\mathrm{pump}}$
and the measured frequency $f_{\mathrm{SA}}%
$, centralized on the pump frequency, which coincides with the
the resonance frequency $f_3=5.74\mathrm{GHz}$
$\left(  f_{\mathrm{SA}}^{\mathrm{c}}=f_{\mathrm{SA}}-f_3\right)$.
Panel $(\mathrm{b}%
)$, subplots $(i),(ii),(iv),(vi)$, plot the same measurement at four different pump powers,
corresponding to $(i)$ the lower power threshold, $(ii)$ and $(iv)$ powers that are in the range of regular SM,
and $(vi)$ the upper power threshold. The red curve in subplot $(ii)$, which has a Lorenzian shape, shows the spectral
density of the SM sidebands as predicted theoretically by the model (Eq. 26 of Ref. \cite
{segev06c}).
The solid red curve in subplot $(i)$ was obtained by numerically integrating the
equations of motion of the model with a nonvanishing noise at the first power threshold and evaluating the spectral density
\cite{segev06c}%
. Subplots $(iii)$ and $(v)$ were obtained by numerically integrating the
equations of motion of the model for the noiseless case, at  input powers corresponding to
subplots $(ii)$ and $(iv)$ respectively, and calculating the spectral density.}%
\label{SM_Phenomenon}%
\end{figure*}
Fig. \ref{SM_Phenomenon}, panels $(\mathrm{a})$ and $(\mathrm{b})$ show
typical experimental results of the SM phenomenon in the frequency domain. The
resonator is stimulated with a monochromatic tone at the resonance frequency
of the third mode $f_{3}$, and the dependence of the SM on the injected pump
power is investigated. At relatively low and relatively high input power
ranges ($P_{\mathrm{pump}}\lesssim-33.25~$dBm $\cup$ $P_{\mathrm{pump}%
}>-23.7~$dBm) the response of the resonator is linear, namely, the reflected
power off the resonator contains a single spectral component at the frequency
of the stimulating pump tone $\omega_{\mathrm{p}}$. In between these power
ranges, regular SM of the reflected power off the resonator occurs (see panel
$(\mathrm{b})$, subplots $(ii,iv)$). It is realized by rather strong and sharp
sidebands, which extend over several hundred megahertz at both sides of the
resonance frequency. The SM frequency, which is defined as the frequency
difference between the pump and the primary sideband, increases as the pump
power increases. The regular SM starts and ends at two power thresholds,
referred to as the lower and the upper power thresholds. The lower power
threshold (panel $(\mathrm{b})$, $i$) spreads over a very narrow power range
of approximately $10%
\operatorname{nW}%
$, during which the resonator experiences a strong amplification of the noise
floor (noise rise) over a rather large frequency band, especially around the
resonance frequency itself. This noise rise can be explained in terms of
nonlinear dynamics theory, as it predicts the occurrence of strong noise
amplification near a threshold of instability
\cite{bifAmp_Wiesenfeld85,BifAmp_Kravtsov01}. The upper power threshold (panel
$(\mathrm{b})$, $vi$) spreads over a slightly larger power range than the
lower one and has similar, but less extreme characteristics.%
\begin{figure}
[ptb]
\begin{center}
\includegraphics[
height=5.7316cm,
width=8.3612cm
]%
{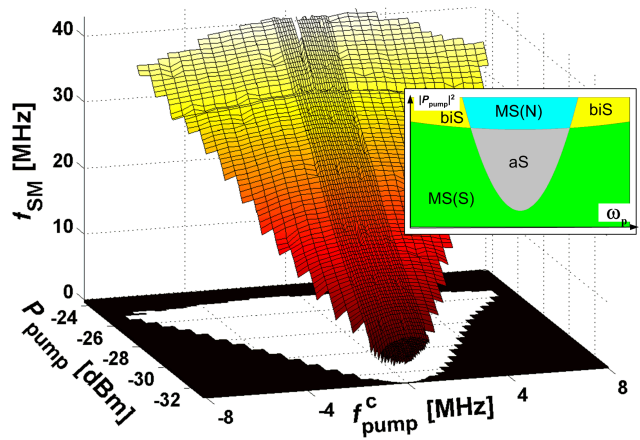}%
\caption{SM frequency $f_{\mathrm{sm}}$ as a function of the pump power and
the pump frequency, centralized on the resonance frequency $f_{\mathrm{pump}%
}^{\mathrm{c}}=f_{\mathrm{pump}}-f_{3}$. The inset shows a schematic diagram
of the stability zones of the resonator as a function of the pump power and
frequency. The green, pale blue, yellow, and grey colors represent the SC
monostable (MS(S)), NC monostable (MS(N)), bistable (biS), and astable (aS)
zones respectively.}%
\label{SM_Freq}%
\end{center}
\end{figure}

The SM nonlinearity is also strongly dependent on the pump frequency. This
dependence is shown in the supplementary video 1, where each frame is a graph
similar to Fig. \ref{SM_Phenomenon}, panel$(\mathrm{a})$, and corresponds to a
different pump frequency, starting at a red shifted pump frequency ($\Delta
f=f_{\mathrm{pump}}-f_{3}<0$) and ending at a blue shifted pump frequency
\ ($\Delta f>0$). Fig. \ref{SM_Freq} plots the SM frequency as a function of
the pump frequency and power. The inset plots the stability diagram of the
resonator (to be discussed below) as a function of the same parameters. The SM
occurs only within a well-defined frequency range around the resonance
frequency. A small change in the pump frequency can abruptly ignite or quench
the SM. Once started though, the modulation frequency has a weak dependence on
the pump frequency. The maximum SM frequency measured with this device is
approximately $41%
\operatorname{MHz}%
$, whereas higher maximum of approximately $57%
\operatorname{MHz}%
$ has been measured with other devices \cite{segev06c}.

We propose a theoretical model according to which the SM originates from a
thermal instability in the microbridge section of the SC stripline resonator.
Current-carrying superconductors are known to have two or more metastable
phases sustained by Joule self-heating
\cite{SM_Self-heatingNormalMetalsSuperconduct}. One is the SC phase and the
other is an electrothermal local phase, known as hotspot, which is basically
an island of normal conducting (NC) domain, with a temperature above the
critical one, surrounded by a SC domain. This phenomenon can be explained by
the heat balance equation holding at more than one temperature. A perturbation
can trigger or suppress the formation of a hotspot and thus the microbridge
can oscillate between instable phases. Such oscillations were often observed
in experiments, for the case of a SC microbridge driven by an external dc
voltage (see review \cite{SM_Self-heatingNormalMetalsSuperconduct} and
references therein).

In the current case, as the microbridge is integrated into a stripline
resonator, the system is driven into instability via externally injected
microwave pump tone. We herein briefly present the corresponding theoretical
model, whereas the full derivation of the equations, as well as experimental
justifications to several assumptions are included elsewhere \cite{segev06c}.
Consider a resonator driven by a weakly coupled feedline carrying an incident
coherent tone $b^{\mathrm{in}}e^{-i\omega_{\mathrm{p}}t}$, where
$b^{\mathrm{in}}$ is constant complex amplitude ($\left\vert b^{\mathrm{in}%
}\right\vert ^{2}\propto P_{\mathrm{pump}})$ and $\omega_{\mathrm{p}}$ is the
driving angular frequency.\ The mode amplitude inside the resonator can be
written as $Be^{-i\omega_{\mathrm{p}}t}$, where $B\left(  t\right)  $ is a
complex amplitude which is assumed to vary slowly on a time scale of
$1/\omega_{\mathrm{p}}$. \ In this approximation, the equation of motion of
$B$ reads \cite{Squeezing_Yurke05}.%
\begin{equation}
\frac{\mathrm{d}B}{\mathrm{d}t}=\left[  i\left(  \omega_{\mathrm{p}}%
-\omega_{0}\right)  -\gamma\right]  B-i\sqrt{2\gamma_{1}}b^{\mathrm{in}%
}+c^{\mathrm{in}},\label{dB/dt}%
\end{equation}
where $\omega_{0}\left(  T\right)  $ is the temperature dependant angular
resonance frequency, $T$ is the temperature of the hotspot, $\gamma=\gamma
_{1}+\gamma_{2}$, where $\gamma_{1}$ is the coupling constant between the
resonator and the feedline, and $\gamma_{2}\left(  T\right)  $ is the
temperature dependant damping rate of the mode.\ The term $c^{\mathrm{in}}$
represents an input Gaussian noise with a zero-mean and a random phase. We
consider a case where the nonlinearity originates by a local hotspot in the
microbridge. If the hotspot is assumed to be sufficiently small, its
temperature $T$ can be considered as homogeneous. The temperature of other
parts of the resonator is assumed to be equal to that of the coolant $T_{0}$.
The power $Q$ heating up the hotspot is given by $Q=\kappa Q_{\mathrm{t}}$
where $Q_{t}=\hslash\omega_{0}2\gamma_{2}\left\vert B\right\vert ^{2}$ is the
total power dissipated in the resonator, and $0\leqslant\kappa\leqslant1$
represents the portion of the dissipated power that is being absorbed by the
microbridge. The heat balance equation reads%
\begin{equation}
C\frac{\mathrm{d}T}{\mathrm{d}t}=Q-W,\label{dT/dt}%
\end{equation}
where $C$ is the thermal heat capacity, $W=H\left(  T-T_{0}\right)  $ is the
heat transfer power to the coolant, and $H$ is the heat transfer coefficient.

Nonlinearity, according to our simple theoretical model, results by\ coupling
the equation of motion of the mode amplitude in the resonator, Eq.
(\ref{dB/dt}), to the thermal balance equation in the microbridge, Eq.
(\ref{dT/dt}). \ The coupling mechanism is based on the dependence of both the
resonance frequency $\omega_{0}$ and the damping rate $\gamma_{2}$ of the
driven mode on the resistance of the microbridge \cite{supRes_Saeedkia05},
which in turn depends on the temperature of the microbridge. We assume the
simplest case, where this dependence is a step function that occurs at the
critical temperature $T_{\mathrm{c}}$. We further assume that all other
parameters are temperature independent. In our previous publication
\cite{Segev06a} we have thoroughly\ investigated the dependence of the
resonance modes on the resistance of the microbridge. We have shown that a
resonance frequency can be smoothly shifted by gradually controlling the
bridge resistance. In addition the damping rate increases up to a certain
maximum, as the resistance increases, but then decreases back to approximately
its original value, when the resistance is further increased.

Due to the step-like dependence of the resistance, and the dependence of the
heat generation, on the temperature of the microbridge
\cite{SM_Self-heatingNormalMetalsSuperconduct}, the system may have in general
up to two locally stable steady states, corresponding to the SC and NC phases.
These states are associated with stationary solutions of the uncoupled Eq. (1)
and (2), where the noise is disregarded.\ The stability of each of these
phases depends on both the power and frequency parameters of the injected pump
tone. In general there exist four different stability zones (see inset in Fig
\ref{SM_Freq}) \cite{segev06c}. Two are monostable zones, where either the SC
phase or the NC phase is locally stable. Another is a bistable zone, where
both phases are locally stable \cite{Baleegh_bifurcation,Baleegh06a}. The
third is an astable zone, where none of the phases are locally stable, and
consequently the resonator is expected to oscillate between the two phases. As
the two phases significantly differ in their reflection coefficients, the
oscillations translate into a modulation of the reflected pump tone. The
mechanism introduced in this theoretical model is somewhat similar to one of
the mechanisms which cause self oscillations in an optical parametric
oscillator \cite{OPO_Suret00}.%
\begin{figure}
[ptb]
\begin{center}
\includegraphics[
height=2.4799in,
width=3.3723in
]%
{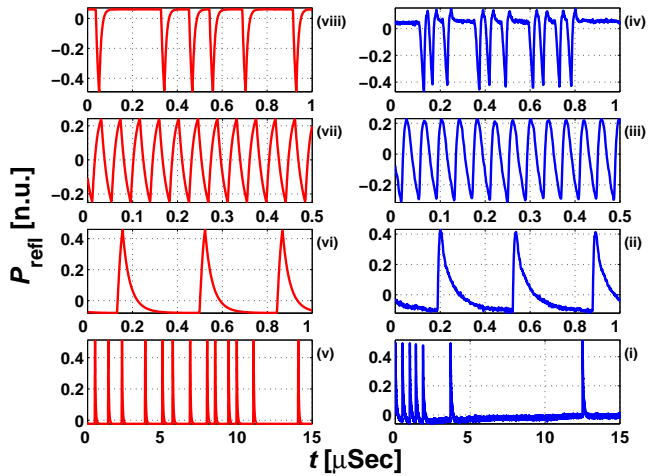}%
\caption{Panels $(i)-(iv)$ show typical experimental results of the SM
phenomenon in the time domain. The reflected power is normalized by the value
of the maximum peak to peak amplitude. AC coupling is employed in the
measurement and thus a zero value represents the average measured power. Each
panel refers to a different pump power range, $(i)$ lower power threshold,
$(ii)$ and $(iii)$ powers corresponding to regular SM. $(iv)$ upper power
threshold. Panels $(v)-(viii)$ show numerical integration of Eqs. (1) and (2)
calculated for the above-mentioned four cases. The graphs are centralized
around their mean value and normalized by the value of the maximum peak to
peak amplitude. In the simulation of the lower and upper thresholds (panels
$(v)$ and $(viii)$), a Gaussian thermal noise, corresponding to
$4.2\operatorname{K}$ and $15\operatorname{K}$ temperatures was assumed
respectively , whereas in the simulation of the regular SM, (panels $(vi)$ and
$(vii)$), the noise was disregarded. Supplementary video 2 shows the SM in the
time domain while gradually increasing the pump power from frame to frame.
Each frame has a subplot similar to panels $(i)-$ $(iv)$, \ and a subplot
similar to \ref{SM_Phenomenon}$(b)$ panels $(ii)$, $(iv)$.}%
\label{simAndExpSMvsTime}%
\end{center}
\end{figure}

The dependence of the SM on the input power, as observed in the time domain
(Fig. \ref{simAndExpSMvsTime}), gives a complimentary understanding of the SM
phenomenon. Below the lower power threshold the resonator is in the SC low
reflective phase, where only a small portion of the injected power is
reflected off the resonator. Once the pump power approaches the power
threshold, sporadic spikes in the reflected pump power occur, as seen in
subplot $(i)$. Near the threshold the device is in sub-critical conditions and
these spikes are caused by a stochastic noise which triggers transitions of
the microbridge from the SC phase to the NC high reflective phase. During a
transition the stored energy in the resonator is quickly discharged and
dissipated. Consequently, after a quick and short worm-up the microbridge
cools down until eventually it switches back to the SC phase. As a result, the
damping rate decreases back to its original value and thus the stored energy
in the resonator is slowly built up again. Once a spike is triggered, the
noise has a negligible effect on the dynamics of the energy discharge and
buildup and therefore the line-shapes of the various spikes are similar. This
behavior can also be clearly observed in the simulation results, shown in
subplot $(v)$.

Regular SM of the reflected power (Fig. \ref{simAndExpSMvsTime} $(ii,vi)$)
occurs when the pump power is set above the lower power threshold. In this
case the pump tone drives the resonator to the astable state and the noise has
a negligible influence. The dynamics of the oscillations is similar to the one
just described where the energy buildup time is strongly dependant on the pump
power. Thus, the oscillation frequency is faster for higher pump power values
(Fig. \ref{simAndExpSMvsTime} $(iii,vii)$). The upper power threshold (Fig.
\ref{simAndExpSMvsTime} $(vi)$) resembles the lower one, but the SC and NC
phases exchange roles. The resonator is in the NC high reflective phase, and
noise-induced spikes temporarily drive it to the SC low reflective phase. The
internal thermal noise in the upper threshold is stronger than in the lower
one and consequently this power threshold range is wider.

In conclusion, we report on a novel nonlinear phenomenon where SM is generated
in a SC microwave stripline resonator. This behavior is robust and occurs in
all of our devices and at various resonance frequencies. A theoretical model
according to which the SM originates by a thermal instability is introduced.
In spite of its simplicity, the model exhibits a good quantitative agreement
with the experimental results. In our measurements we find strong and
correlated nonlinear phenomena accompanying the SM \cite{segev06d}. One of
these phenomena is self-excitation (SE) of coupled resonance modes, which is
shortly described in a supplementary figure. This phenomenon may be exploited
for state readout in quantum data processing systems \cite{SQZ_Buks05}.

\textbf{Acknowledgements} We thank Bernard Yurke, Ron Lifshitz, Mile Cross,
Oded Gottlieb, and Steven Shaw for valuable discussions. This work was
supported by the German Israel Foundation, the Israel Science Foundation, the
Deborah Foundation, the Poznanski Foundation, and MAFAT.

\textbf{Author Information} Reprints and permissions information is available
at npg.nature.com/reprintsandpermissions. The authors declare no competing
financial interests. Correspondence and requests for materials should be
addressed to S.E.\symbol{126}(email: segeve@tx.technion.ac.il).

\bibliographystyle{naturemag}
\bibliography{Bibilography}

\end{document}